# Mode hitching in traveling-wave optical parametric amplification

Joseph Kelly,[1,2,*] Eleanor Fradgley,[1] Lida Zhang (张理达),[3] and Vincent Boyer[1,†]

[1]*School of Physics and Astronomy, University of Birmingham, Edgbaston, Birmingham B15 2TT, United Kingdom*
[2]*Department of Physics, King's College London, Strand, London WC2R 2LS, United Kingdom*
[3]*School of Physics, East University of Science and Technology, Shanghai 200237, China*

Optical parametric amplifiers (OPAs) in traveling-wave configuration can generate localized spatial quantum correlations between a signal and an idler beam, a useful resource for quantum imaging. To characterize the spatial correspondence between the optical modes of the signal and those of the idler, we look at the classical transverse dynamics of these beams when they propagate in a generic thick OPA at a nominally small angle. In these conditions, the beams tend to copropagate while maintaining a fixed separation, a phenomenon that we term hitching. We develop a model for hitching, validated by a numerical simulation, and provide an experimental demonstration using four-wave mixing in a hot atomic vapor. They show that the OPA gain is the primary influence on the final hitching distance. These results have implications for the generation of multi-spatial-mode squeezed light for quantum imaging applications, where the exact spatial correlations between the quantum fluctuations of the signal and the idler are of prime importance.

## I. INTRODUCTION

Optical parametric amplification is at the heart of quantum-state production [1,2]. In an optical parametric amplifier (OPA), a signal optical field is amplified simultaneously with the creation of an idler field, through nonlinear coupling with one or more additional pumping fields. The coherence of the process ensures that signal and idler fields, here referred to as twin beams, are created and correlated in a shared quantum state. This can be used to create photon pairs that are entangled in polarization [3], position, and momentum [4] or fields that are entangled or squeezed in quadrature [5] or squeezed in intensity difference [6].

In quantum imaging, one typically seeks to engineer squeezing or entanglement independently at any position in a transverse section of the optical field [7–10]. This can be done with a traveling-wave amplifier that relies on a thin nonlinear medium. The medium naturally couples the signal and idler fields locally, creating spatially localized quantum correlations. An equivalent statement is that the amplifier couples arbitrarily narrow modes of the signal and the idler at any position of the thin medium.

Having a thin medium removes the more complicated dynamics associated with propagation [11,12]. Specifically, twin beams can overlap more easily along the full length of a thin medium, even when propagating in different directions. In contrast, propagation through a thick medium could cause the correlated fields to separate spatially while being created. In practice, the medium must always have a finite length to generate a finite amplification gain. Typically, the production of quadrature squeezing and entanglement requires a medium that is much longer than the wavelength of the light. For a correspondingly thick medium, even if the fields are strictly copropagating, diffraction puts a lower limit on the width of the signal and idler beams that can be coupled, leading to the emergence of a coherence area [9,11]. In this work, we go one step further and look at the impact of phase conjugation on the joint propagation of the signal and idler modes, as well as the possible modification of the coherence area.

The issue of propagation arises when the signal and idler beams propagate with a small angle between them. This is a common situation: When pumping the medium with one or several plane waves along $z$, the conservation of the total transverse momentum, which is null, makes the propagation of the twin beams symmetric with respect to the $z$ direction. This becomes relevant when the amplification is phase matched for a finite angle between the twin beams and the pumping direction. It raises the question of what happens when initially narrow twin beams, with a size as small as the coherence area, separate while propagating in the gain medium. Since they stimulate the nonlinear process and get amplified while generating their partner beam along their separate trajectories, the expectation is the production of wide-area signal and idler fields. We show here that for a small angle between the signal and the idler, this is not the case. The signal and idler effectively copropagate, in apparent violation of the phase-matching condition, at a fixed distance from each other. In the following, we refer to this phenomenon as hitching. As long as the width of the seed beam is larger

*Contact author: joseph.2.kelly@kcl.ac.uk
†Contact author: v.boyer@bham.ac.uk

than or equal to the coherence length, the width of the twin beams stays roughly constant.

The effect described here is separate from the walkoff effect observed in type-II parametric down-conversion due to the birefringence of the nonlinear medium. In this case, a mismatch between the Poynting vector and the wave vector forces the signal to walk off the pump beam [3,13]. The effect presented here may be smaller in magnitude and occurs even in isotropic media. It would likely be masked by the transverse walkoff in a typical birefringent crystal.

Paraxial propagation of twin beams in an OPA has previously been considered in the context of diffraction and absorption control [14,15]. Cancellation of diffraction on the signal beam is achieved when the angular dispersion of the beam is approximately zero, that is to say, the accumulated phase of a plane wave on traversal of the amplifying medium does not depend on the transverse wave vector. This effect may be realized for specific values of the complex direct and cross susceptibilities of the medium for the twin beams [15]. The problem we are considering here is different. Starting from OPA parameters which are known to produce highly quantum states of light, namely, low absorption on both twin beams and high gain, we look at the evolution of the average position of the beams during propagation in the medium and show that the apparent trajectory of the beams upon amplification deviates from a straight line.

The paper is organized as follows. Section II introduces a basic theoretical model of the hitching effect, which is then solved numerically. This allows us to extract the main physical characteristics of the phenomenon. Section III describes an implementation of OPA based on four-wave mixing in a hot atomic vapor, and the experimental setup that allows us to measure the hitching effect. Section IV presents our experimental results and shows how the hitching distance depends simply on the amplification gain. We conclude in Sec. V by looking at the implications of these results on the generation of multi-spatial-mode squeezed light for quantum imaging.

## II. MODEL

The basic physics of the hitching effect is simple. Parametric amplification is stimulated by both the signal and the idler. As a consequence, it occurs with a higher gain in regions where both twin beams are present, i.e., where the signal and the idler overlap. While the phase-matching condition pulls the twin beam apart, preferential amplification in their overlap region keeps their average separation bounded so that they propagate in lockstep.

In this section, we look at the minimum theoretical model of twin-beam generation that shows transverse hitching and its dependence on experimental parameters. To this effect, we consider a pair of quasidegenerate coupled beams of unspecified polarization, created by optical parametric amplification, here referred to as modes 1 and 2. This is typically produced by down-conversion or four-wave mixing in a traveling-wave configuration. However, we keep it generic at this stage. To rule out transverse walkoff effects due to birefringence, we only consider an isotropic medium with a scalar index of refraction, possibly complex to represent absorption. Linear or nonlinear refraction effects are also annulled by making the medium spatially homogeneous, a condition that would be experimentally achieved by using wide and collimated pumping beams. This condition also ensures that there is no transverse momentum associated with the pumping. Since we are interested in walkoff effects between the coupled beams, we assume them to be copropagating to a first approximation, possibly with a small angle between them. In the paraxial approximation, their respective electrical field amplitudes can be written as

$$E_1(x,z,t) = \mathcal{E}_1(x,z)e^{i(k_1 z - \omega_1 t)}, \quad (1)$$

$$E_2(x,z,t) = \mathcal{E}_2(x,z)e^{i(k_2 z - \omega_2 t)}. \quad (2)$$

To simplify the problem, we assume that the beams have similar frequencies $\omega_1 \simeq \omega_2$ and therefore similar wave vectors $k_1 \simeq k_2 = k$. The beams propagate along the $z$ direction in a nonlinear medium placed between $z = 0$ and $z = L$, and we only consider the $x$ transverse direction. In real space, $\mathcal{E}_1(x,z)$ and $\mathcal{E}_2^*(x,z)$ interact locally via the cross susceptibility $\chi$. In the reciprocal transverse space, $\chi$ couples the slowly varying envelopes $\mathcal{E}_1(k_x, z)$ and $\mathcal{E}_2^*(-k_x, z)$, which propagate according to

$$\frac{\partial}{\partial z}\begin{bmatrix}\mathcal{E}_1 \\ \mathcal{E}_2^*\end{bmatrix} = i\begin{pmatrix}-\Delta k + a_1 & b \\ -b & \Delta k - a_2^*\end{pmatrix}\begin{bmatrix}\mathcal{E}_1 \\ \mathcal{E}_2^*\end{bmatrix}, \quad (3)$$

where $\Delta k = \frac{k_x^2}{2k}$ is the paraxial phase responsible for diffraction. The coefficients $a_1$ and $a_2$ are proportional to the linear susceptibilities for modes 1 and 2 and account for single-beam-propagation effects, such as absorption and refraction. The cross coupling is given by $b = \frac{k}{2}\chi$ and has to be real for efficient amplification. The amplification couples mode 1 transverse momentum $k_x$ with mode 2 transverse momentum $-k_x$. When the process is stimulated by a seed field $\mathcal{E}_s(k_x)$ in mode 1 at position $z = 0$, the solutions to Eq. (3) inside the medium are

$$\mathcal{E}_1(k_x, z) = \mathcal{E}_s(k_x)e^{i\delta a z}\left(\cosh \xi z + \frac{ia}{\xi}\sinh \xi z\right), \quad (4)$$

$$\mathcal{E}_2^*(k_x, z) = -\mathcal{E}_s(-k_x)e^{i\delta a z}\frac{ib}{\xi}\sinh \xi z, \quad (5)$$

where

$$\delta a = \tfrac{1}{2}(a_1 - a_2^*), \quad (6)$$

$$a = \tfrac{1}{2}(a_1 + a_2^*) - \Delta k, \quad (7)$$

$$\xi = \sqrt{b^2 - a^2}. \quad (8)$$

This set of equations allows us to propagate both beams in reciprocal transverse space.

We first consider the case of a fully transparent medium, which is achieved when $a_1$ and $a_2$ are real. In the absence of refraction, that is, when $a_1 = a_2 = 0$, the nonlinear process is phase matched for fully copropagating beams. In other words, the gain, defined by the ratio $G = |\mathcal{E}_1(k_x, L)|^2/|\mathcal{E}_s(k_x)|^2$, is maximum for $k_x = 0$. In the context of imaging, it is advantageous to be able to spatially separate the two beams. It can be achieved in the far field when they propagate at a small angle in the $x$-$z$ plane. To make the amplifier efficient in this geometry, we introduce a small angle between the phase-matched directions by adding some refraction on the two beams, that is, by giving finite real values to $a_1$ and $a_2$. From Eqs. (4)–(8) it is clear that it does not matter how the refraction is distributed between the modes.

Solving the propagation of beams of finite transverse sizes is achieved by propagating their plane-wave components. More specifically, the envelope of the seed beam is Fourier decomposed at $z = 0$ as

$$\mathcal{E}_s(k_x) = \frac{1}{\sqrt{\pi}} \int \mathcal{E}_s(x) e^{-ik_x x} dx, \qquad (9)$$

where the envelope $\mathcal{E}_s(x)$ represents the seed field profile, with an added phase gradient accounting for the finite angle of incidence. After propagation, the slowly varying envelopes of modes 1 and 2 are

$$\mathcal{E}_i(x, z) = \frac{1}{\sqrt{\pi}} \int \mathcal{E}_i(k_x, z) e^{ik_x x} dk_x, \quad i = 1, 2, \qquad (10)$$

where the $\mathcal{E}_i(k_x, z)$ are given by Eqs. (4) and (5).

Figure 1 shows how such beams propagate in a finite-length medium, when the optical amplifier is seeded with a Gaussian beam in mode 1. The width of the beam is chosen to be small and remains small throughout the length of the medium, i.e., it is commensurate with the coherence length. For reference, the propagation of the seed in free space is shown in Fig. 1(a). As expected, mode 1 is amplified [Fig. 1(c)] and mode 2 is created and amplified [Fig. 1(d)]. Figures 1(e) and 1(f) show these respective modes normalized to a fixed power at each position $z$, to better appreciate the apparent trajectory. The conservation of transverse momentum, embodied by the phase-matching condition, is visible on the beams emerging from the medium, symmetrically positioned on each side of the $z$ axis with mode 1 propagating in the same direction as the seed.

What happens inside the medium is more complex. First, the width of the beams does not increase as much as one would expect in a naive model where mode 2 would appear as if born from every point on the mode 1 trajectory, and reciprocally. This would result in modes 1 and 2 being overlapped at all $z$ and having a width that increases linearly with $z$.

Second, the apparent direction of propagation for both beams is different from the directions seen before (for mode 1) and after (for both modes 1 and 2) the medium. Mode 2 propagates along the $z$ direction exactly, while mode 1 is drawn towards the $z$ direction, so that the distance between the two beams does not get larger than their size. As a result, mode 1 emerges from the medium at a position that is shifted with respect to free-space propagation and mode 2 always emerges close to the seed (mode 1) position on the medium input interface. Figure 1 shows the effect for an amplification factor of mode 1 power of 30.

Note that light still propagates along the original direction at all times. Figures 1(e) and 1(f) show the average wave vectors $(\langle k_x \rangle, q)$ of modes 1 and 2 at various positions along the propagation. The directions of the wave vectors remain constant, as long as the relative values of the seed average angle and the indices of refraction $a_1$ and $a_2$ are chosen so that the amplification is phase matched. The effect of deflection is given by the fact that the beams are preferentially amplified in the region where they overlap, which keeps their transverse intensity (and field) center of mass in proximity. In terms of coupled modes emerging from the medium, the hitching effect is real and should have consequences on the structure of spatial squeezing and quantum correlations on the output of a thick OPA.

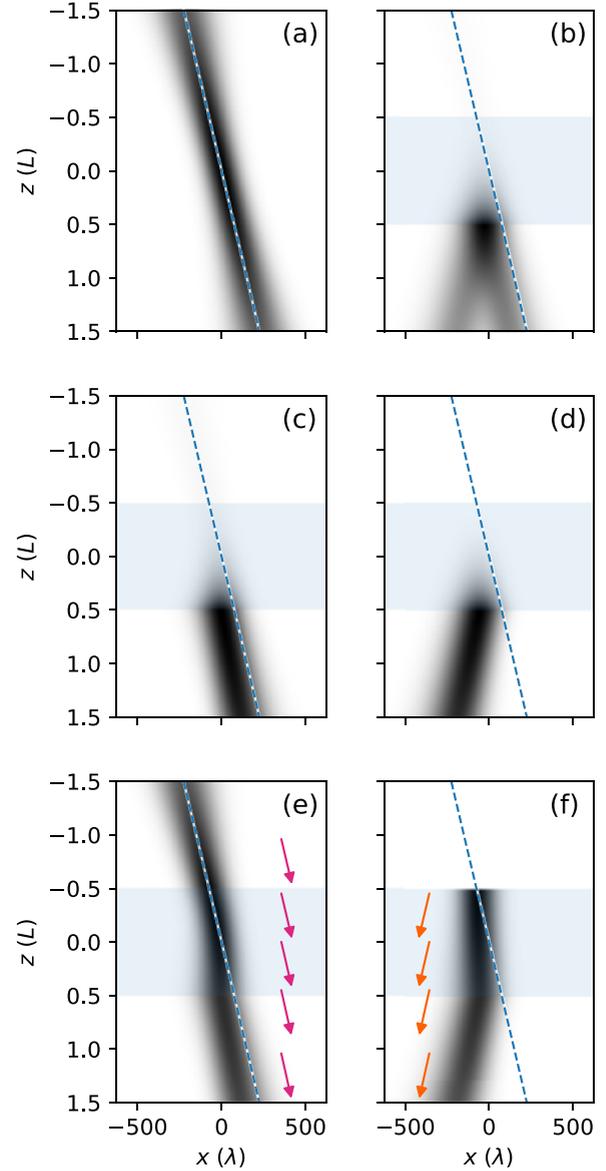

FIG. 1. Numerical simulation showing beam propagation and amplification in a multimode OPA. Each panel represents the intensity distribution on a white (no light) to black (maximum intensity) scale, as a function of the propagation direction $z$ and the transverse direction $x$. The amplifier is seeded on mode 1. This mode is shown in (a), propagating in free space (no medium). It is a Gaussian beam inclined at 3 mrad with respect to the $z$ axis, with an electric-field standard deviation $\sigma = 100\lambda$, where $\lambda$ is the wavelength. The white and blue dashed line, present in all panels, is a guide to the direction of the free propagating seed. Panels (c) and (d) show separate mode 1 and mode 2 amplification and propagation, respectively, when the amplifying medium, show in light blue, is present. The medium has a length $L = 5 \times 10^4 \lambda$ and the medium parameters appearing in Eq. (3) are $a_1 = a_2 = 2.8 \times 10^{-5}$ rad/$\lambda$ and $b = 1.0 \times 10^{-4}$ rad/$\lambda$. Panel (b) shows both modes 1 and 2 simultaneously, as they are in a real amplifier. Panels (e) and (f) show the same data as in (c) and (d), where the intensities are normalized to the maximum intensity for each mode at each $z$ position. The magenta and orange arrows represent the average wave vectors, in arbitrary units, of the modes in and outside the medium, at the $z$ positions where the arrows originate. The large gain, 30, makes it impossible to distinguish the seed in (c).

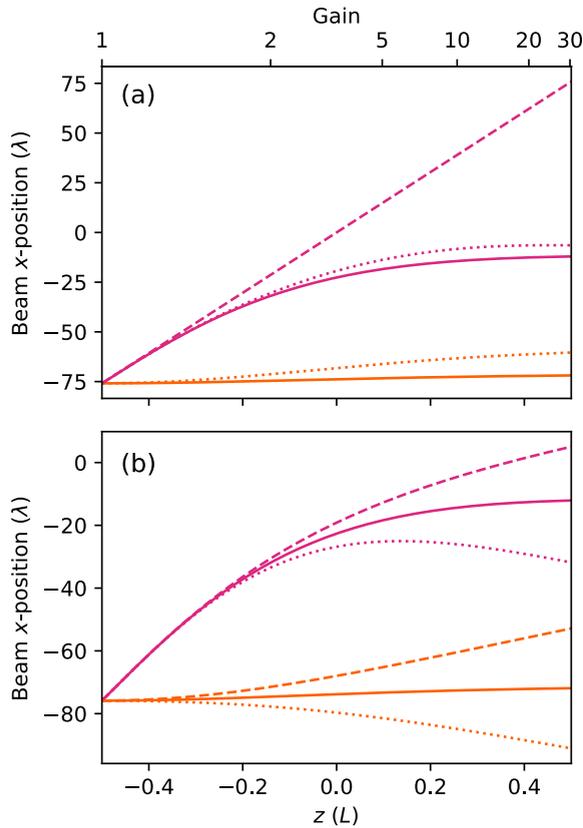

FIG. 2. Numerical results showing the evolution of the transverse position of the intensity center of mass of mode 1 (magenta lines) and mode 2 (orange lines), as a function of the distance of propagation in the medium. In (a) there is no absorption in the medium (the imaginary parts of $a_1$ and $a_2$ are zero). The dotted lines are the positions of the maximum intensity of the modes. The dashed line shows the mode 1 free-space trajectory. The top axis shows the gain on mode 1 total power, for (a) only. In (b) we introduce absorption on the twin beams. The solid lines are the same as in (a). The dotted and dashed lines correspond to absorption on modes 1 and 2, i.e., with a finite imaginary part for $a_1$ and $a_2$, respectively.

Although the coupled modes lock to each other when propagating, their positions when emerging from the medium do not coincide exactly. In Fig. 2 we show the apparent trajectory of the modes across the medium in the conditions of Fig. 1, as measured by the center of mass of their intensity profile. The plot shows that mode 2 is initially created at the position of mode 1. During propagation, the modes pull away from each other until the distance between them stabilizes as the gain on the seeded mode exceeds 2. This is reminiscent of the propagation of matched pulses observed in an OPA [16], where coupled pulses with different group velocities get hitched to each other longitudinally. Here the equivalent to the group velocity is the transverse velocity of the pulses, as measured by the position of their intensity center of mass or, alternatively, maximum intensity. In the geometry of our model, the transverse velocity of mode 2 remains roughly zero, while the transverse velocity of mode 1, initially equal to the free-space velocity, drops to zero to match the velocity of mode 2.

If we introduce loss (absorption) into one of the modes, the situation is modified as in Fig. 2(b). The separation between the modes becomes hitched quantitatively as in the lossless case, but the absolute positions of the modes tend towards the nominal propagation direction of the lossless mode. Note that the net gain at the output of the amplifier is reduced when there is loss. This means that for a given net amplifier gain, hitching will occur after a shorter propagation distance in the medium in the lossy case, compared to the lossless case.

Hitching is a two-beam effect. The trajectory of each of the beams is a function of the position of the other beam and it is not necessarily a straight line. This is in contrast to refraction, which can deflect beams instantly at interfaces. This refraction, observed for a single beam in an atomic vapor subject to electromagnetically induced transparency and coherence diffusion, has been used to control diffraction [14]. In this case, the direction of propagation, dictated by loss rather than gain, does not depend on the longitudinal position inside the medium. The dynamics observed here is more related to predicted diffraction control using four-wave mixing (4WM) in an atomic vapor [15], in that the whole propagation through the medium must be considered to account for the position and size of twin beams at the output of the medium.

In the following sections, we implement an OPA based on 4-wave mixing in an atomic vapor that displays the generic hitching feature discussed above. In this setup, the length of the amplifier, $L$, is fixed and cannot be varied. Instead, experimental conditions are altered to achieve a range of gains and losses. The transverse positions of mode 1 and mode 2 are measured after propagating by length $L$ to the output facet of the amplifier. At low gain and low loss, the modes are unhitched and mode 1 emerges at a position corresponding to free propagation throughout the amplifier. The generated mode, mode 2, stays in position. At high gain, the modes are hitched and they emerge locked transversely closer to each other.

## III. EXPERIMENTAL SETUP

We implement a traveling-wave OPA using the so-called double-lambda scheme in a $^{85}$Rb vapor [17,18]. A collimated cw pump beam with a total power of 1.2 W is tuned to the blue of the $D_1$ line at 795 nm and couples to sidebands located at plus or minus one ground-state hyperfine splitting (3 GHz) away from the pump frequency, as shown in Fig. 3. Other parameters are broadly similar to those in Ref. [19]. The Rayleigh range of the pump beam is much larger than the scale of the experiment, so the pump has a negligible transverse-momentum distribution. A probe beam, corresponding to mode 1 (the signal), intersects the pump beam at a small angle in a 20-mm-long cell filled with a hot $^{85}$Rb vapor. The probe beam seeds the 4WM and is amplified while generating a beam in mode 2 (the idler), called here the conjugate. This is a parametric process, and the conjugate frequency is uniquely determined by energy conservation. As the probe beam is closer to the atomic resonance, it sees an index of refraction slightly higher than that of the other beams, which is effectively 1. Relating this to the theory section, this means that $a_2$ is negligible, while $a_1$ has a finite value. The latter is potentially complex if there is absorption on the probe beam. As a result, the process is phase matched for an angle of 4–8 mrad between each of the twin beams and the pump direction [9]. The experiment operates in steady state, in the low-pump-depletion regime, in which the evolution of the twin beams is well described by Eq. (3). In these conditions, the process can exhibit a gain on the probe beam of up to 50 in a single pass. Moreover, the cross-coupling coefficient $b$ is mostly real [20], as considered in the model section.

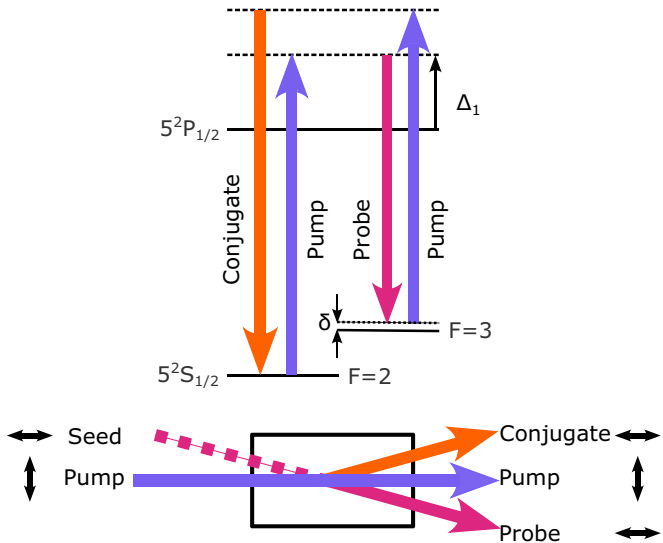

FIG. 3. Shown on top is the double-lambda scheme on the $D_1$ line of $^{85}$Rb and on the bottom the geometry of the 4WM within the rubidium cell. The double-ended arrows next to each beam refer to the orthogonal polarizations of these beams as viewed in the $x$-$y$ plane perpendicular to the propagation direction of the beam.

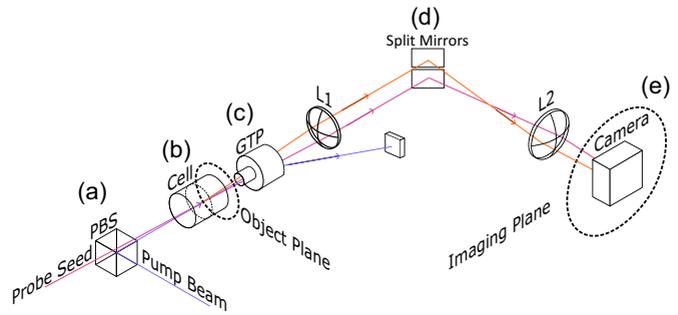

FIG. 4. Experimental setup. Probe (magenta) and pump (purple) beams are superimposed on a polarizing beam splitter (PBS) and overlapped at a small angle in the vapor cell. After the cell, the pump is separated from the twin beams with a Glan-Thompson polarizer (GTP). A couple of lenses L1 and L2 form a $2f$-$2f'$ optical system that images the cell output onto a camera with magnification 3. A split mirror in the Fourier plane independently directs the probe and the conjugate (orange) beams so that their images are separated on the camera.

The relative position of the output beams is determined by imaging the output facet of the vapor cell on a CCD camera. Figure 4 shows how the twin beams are imaged separately so that their transverse positions can be resolved separately. Although we cannot measure the absolute or relative positions of the twin beams, we can measure changes in their positions. This is done by varying the system parameters that affect the gain and loss of the amplifier. One of these parameters is the pump intensity. In order to measure the effect of a range of intensities simultaneously, we position an elongated probe beam such that, in the amplifier, the probe beam extends from the center of the pump beam, where the intensity is maximum, up to the edge of the pump beam, where the intensity is near zero. The profiles of both beams are largely Gaussian. The geometry of the beams is shown in Fig. 5.

The vertical beam width of the probe in the gain medium is 0.5 mm. It is larger than the coherence length but narrow enough to ensure that the pump intensity varies little across the probe in the vertical direction, which is also the direction in which the hitching occurs. The pump diameter is 2 mm.

Images of the twin beams are captured with the probe seed beam both blocked and unblocked so that remnant pump light can be measured and subtracted from the twin-beam images. The images are analyzed in amplitude and vertical position, as shown in Fig. 6. Statistical uncertainties on these measurements are evaluated across groups of five spatially adjacent values. These data are collected as system parameters, such as pump and probe frequencies, are varied to produce a range of values for coefficients $a_1$ and $b$.

## IV. RESULTS AND DISCUSSION

At first, the atomic parameters, pump power and laser beam detunings, have been chosen to give pure gain and no loss on the probe and conjugate. The probe and conjugate positions measured at the output of the vapor cell, as functions of the net probe gain, are shown in Fig. 7. Gain variability is caused either by the intensity profile of the pump or by directly varying the pump power.

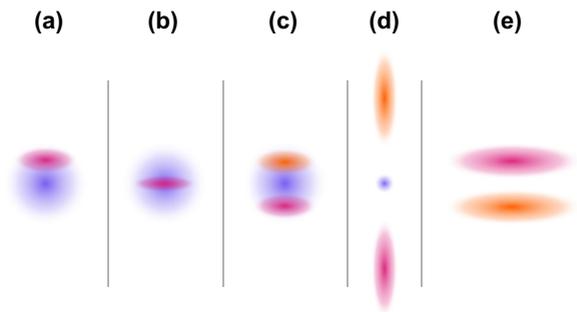

FIG. 5. Diagram of the beam profiles at the locations indicated in Fig. 4. (a) The orthogonally polarized pump and probe seed beams are superimposed at a small angle. Before the cell, the probe is above the pump axis. (b) Both beams are focused and centered on the cell, where 4WM occurs, producing gain on the probe and generating the conjugate beam. (c) After the cell, both twin beams are present, traveling in opposite sides of the pump axis. (d) In the Fourier plane, between lenses L1 and L2, the aspect ratio of the twin beams is inverted. Here most of the pump light has already been removed by the output polarizer and the rest passes between the split mirrors. (e) Finally, the probe and conjugate beams are projected onto the imaging plane with a magnification factor of 3.

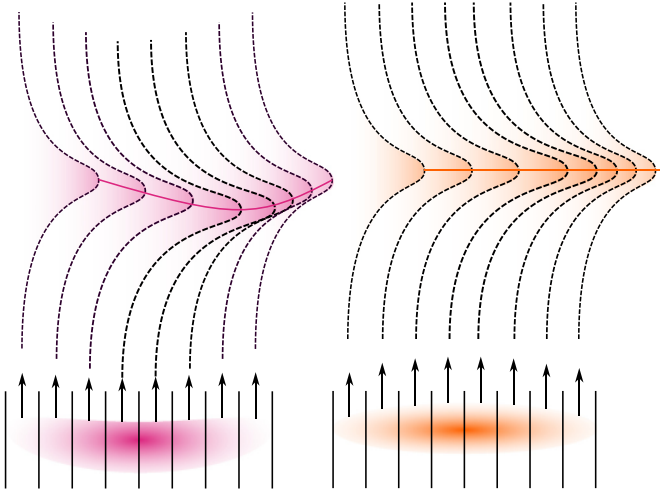

FIG. 6. Diagram of the image data processing pipeline. The images of the probe and conjugate, at the bottom, are analyzed in amplitude and vertical position by Gaussian fitting after averaging over uniform sections across the horizontal direction. In practice, we consider 100 of these segments per image.

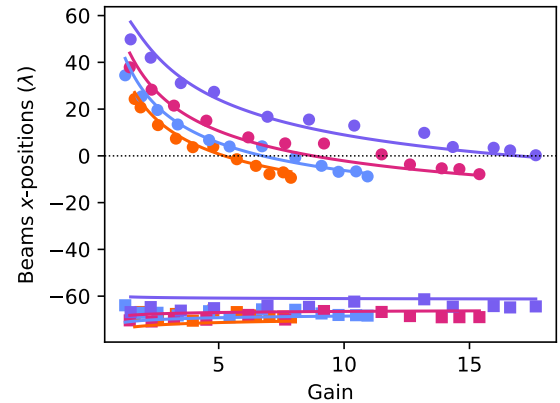

FIG. 8. Positions of the probe (circles) and conjugate (squares) beams at the output of the vapor cell, as a function of the net probe gain, for various levels of probe absorption. The symbols are experimental data. The solid lines are a fit by our theoretical model, with values for the imaginary part of the $a_1$ coefficient of $2.3 \times 10^{-5}$ rad/$\lambda$ (orange), $1.7 \times 10^{-5}$ rad/$\lambda$ (blue), $1.3 \times 10^{-5}$ rad/$\lambda$ (magenta), and 0 (purple).

As expected for a lossless OPA, mode 2 (the conjugate) emerges in an almost constant position, regardless of the gain value. The situation is very different for mode 1 (the probe). At low gain, it emerges at the position that corresponds to free propagation, as indicated by the dashed line in Fig. 2(a). As the output gain increases, the point at which the probe starts hitching to the conjugate, which is roughly when the intermediate gain reaches 2, as shown in that same figure, occurs earlier in the cell. This leads to a hitching distance that decreases as the output OPA gain increases.

The data in Fig. 7 are fitted using Eqs. (4), (5), and (10). As explained above, there is no direct coupling for the conjugate; therefore $a_2 = 0$. Since there is no loss on the probe, the direct coupling $a_1$ is a real parameter, whose value is set so that the process is phase matched at the experimentally determined angle of 5 mrad. The value of the cross-coupling coefficient $b$ is determined from the probe gain measured in each segment. The only free parameters are the vertical positions of the curves, since the optical setup does not allow us to make an absolute determination of the hitching distance. Under this restriction, the experimental results agree remarkably well with the theory curves, indicating that the main features of the hitching effect are captured.

In a second experiment, we tune the optical parameters so that the probe experiences absorption in the atomic medium. This is achieved by reducing the detuning of the probe with the matching atomic transition, denoted by $\Delta_1$ in Fig. 3. The smaller the detuning, the higher the absorption. Figure 8 shows the changes in the probe and conjugate exit positions for different values of the detuning. The data are fitted as previously done, but this time allowing the real part of the $a_1$ coefficient to vary as the detuning changes. It shows that at constant net gain, the distance between the twin beams decreases as the probe loss increases. This is consistent with the theoretical description given in Sec. II. Loss causes earlier hitching in the medium, leading to a reduced hitching distance. The data also show that as the absorption in the probe increases, the propagation of the twin beams becomes biased toward the geometric conjugate propagation direction, as expected from the numerical results in Fig. 2(b).

The experimental results are in good agreement with the theory developed in Sec. II. Before concluding, we should rule out any imperfections or other effects that could potentially result in a similar hitching phenomenon. There are two domains where the experimental conditions depart from the assumptions of the model. They are both linked to a slight dependence of the index of refraction seen by the probe on the atomic and optical parameters of the experiment. The first factor to consider is that the pump beam has a finite extent instead of being a true plane wave. Due to cross-Kerr coupling with the probe beam and to a lesser extent the further-detuned conjugate beam, pump intensity gradients can result in Kerr lensing and deflection of the twin beams when these are not restricted to the center of the pump, where the pump intensity is the most homogeneous. As shown in Fig. 5(b), the twin beams probe the center of the pump in the vertical direction, which is the direction in which we analyze the hitching effect. This precaution also removes the risk of having a differential amplification of the top and bottom parts of the twin beams, due to a vertical gradient of pump intensity.

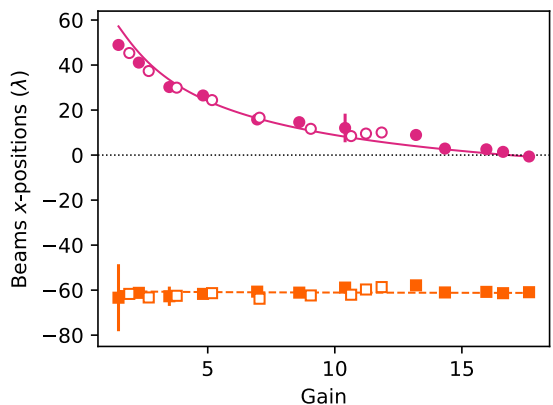

FIG. 7. Positions of the probe (circles) and conjugate (squares) beams at the output of the vapor cell, as a function of the net probe gain at system parameters corresponding to a low-loss case. The points are experimental data. Closed symbols represent data obtained across the pump beam profile, as the gain varies with the local pump intensity. Open symbols represent data derived from the center of the pump beam, when the overall pump power is varied, in a separate experiment. The solid lines represent a fit to the closed symbols by our theoretical model, with values for the imaginary part of the $a_1$ coefficient of 0. The general effect is increased hitching with higher probe gain.

The second effect to take into account is that the phase-matching condition depends on both the pump intensity and the one-photon detuning. A partial loss of phase matching for some values of the gain and the loss, and the resulting preferential amplification of some of the transverse components of the twin beams, could lead to a deflection of these beams. In the experiment, we tuned the angle of the probe with respect to the pump to maximize the gain for average atomic parameters, thereby reducing the effect of changing phase matching during the scans.

These two effects are different. Cross-Kerr lensing operates in real space, whereas the partial loss of phase matching operates in reciprocal space; however, they both affect the positions of the twin beams in the far field rather than in the near field, that is, in the gain region and its immediate output. For safety, we have checked that no discernible deflection of the probe beam is observed in the far field, meters away from the cell, when the pump is turned on. This means that no possible contribution to lateral beam displacement can be attributed to such deflection effects after only 20 mm of propagation in the cell. This conclusion is strengthened by the fact that hitching data obtained at the center of the pumping beam, where there are no intensity gradients, are fully consistent with the data obtained over the full pump beam profile, as shown in Fig. 7.

In general, the absence of probe shifting in the far field is consistent with the picture presented in Fig. 1, where the direction of propagation of the probe is unchanged after amplification, while having an apparent shift inside the gain medium. This is because in our system of a diluted atomic vapor, the amplification effect created by the cross coupling between the twin beams, embodied by the coefficient $b$ in Eq. (3), dominates over the single-beam propagation effects embodied by coefficients $a_1$ and $a_2$, at least over the length of the medium. This also rules out strong self-interaction nonlinear effects such as soliton attraction [21].

Finally, a single curve of the hitching distance was produced as a function of the position within the pump beam, for optical parameters that are optimum for squeezing or entanglement generation [19,22–27]. This is shown in Fig. 9. The distance between the beams is minimal at the center of the pump beam, where the pump intensity and the gain are maximal.

We expect this effect, shown here classically, to persist in the quantum regime. Indeed, in the absence of loss, Eq. (3) can be quantized simply by replacing the classical field amplitudes by their corresponding quantum operators. Thus, twin narrow beams generated by the classical amplifier, such as those exemplified in Fig. 1, are a good representation of modes coupled by the quantum amplifier and, as a result, will be quantum entangled at the output of the gain region. Considering now a wide input probe beam, which could be decomposed into a superposition of many narrow parallel beams, we see that it will generate a wide conjugate beam, which will be a superposition of narrow modes that are the conjugates of the probe constituent narrow modes. As a result, probe and conjugate beams will exhibit localized quantum correlations in their transverse intensity profiles, whose locations will follow the classical hitching dynamics shown in Fig. 9.

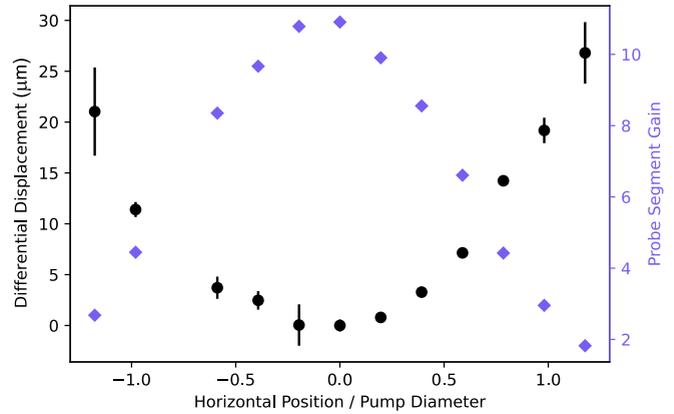

FIG. 9. Experimental data showing the vertical hitching distance between the probe and conjugate segments against horizontal beam position, at system parameters optimal for squeezing. Alongside is the probe gain against horizontal beam position, which varies similarly to the hitching distance but with the opposite sign.

This observation is crucial for quantum imaging experiments using OPAs of finite lengths with inhomogeneous pumping. A typical scenario would be to use quantum-correlated twin beams to perform absorption imaging on a faint object, one beam performing the probing and the other beam serving as an intensity reference [28]. Sub-shot-noise imaging is achieved by subtracting the reference from the probe. This can be done in the near field of the OPA, where the twin beams display spatially identical quantum fluctuations, or in the far field of the OPA, where the quantum fluctuations are symmetrical with respect to the propagation axis. In practice, the spatial correspondence between the fluctuations in the far field is affected by the differential cross-Kerr lensing on the twin beams [29] presented earlier. When imaging in the near field, inhomogeneous gain leads to spatially varying hitching, as described in this paper and shown in Fig. 9.

When using an OPA to generate entanglement or quadrature squeezing, field analysis by means of homodyne detection must take the spatial structure of quantum correlations into account, and this should include the hitching effect. It is noteworthy that in some vapor-cell experiments, such spatial matching has been done by generating the local oscillators of the homodyne detectors using the same amplification process, i.e., using an OPA similar to the one generating the entanglement. This ensures a perfect matching of the spatial modes across the probe and conjugate frequencies, regardless of hitching. The approach has been used to measure the entanglement between the probe and the conjugate spatial modes [26], as well as the quadrature squeezing of a bichromatic field containing components at the probe and conjugate frequencies [27].

## V. CONCLUSION

We have presented a theoretical and experimental study of the propagation of twin beams (signal and idler) in a thick OPA. It shows that under a copropagation arrangement with a small angle between the input probe and a collimated pump (or generally a pumping mechanism that does not introduce spatial phase modulation), the twin beams hitch to each other instead of pulling apart.

This effect is generic in parametric amplification and is distinct from the spatial walkoff observed in birefringent media. It is borne out of the tradeoff between twin-beam production, which occurs where the beams overlap, and the phase-matching condition, which pulls the beams apart. This mechanism kicks in when the OPA gain exceeds 2. The final hitching distance between the beams decreases with the amplifier gain. When one of the twin beams is subject to absorption, the mean final position of the combined beams depends on the relative optical power between them, with the brighter one having the largest influence on their common direction of propagation.

The theory presented in this paper assumes a homogeneous pump of infinite extent. In the more realistic case of a pump of finite width, calculating the structure of the quantum correlations at the output of a low-gain optical amplifier can be done by performing the Schmidt decomposition of the emitted biphoton wave function [30]. The extension of this idea to a higher-gain regime, the condition for the hitching effect to appear, is a supermode decomposition, which reduces the parametric interaction to a collection of independent single-mode amplifiers [31].

We have shown that differential hitching due to inhomogeneous pumping is prevalent in the parameter range where large continuous-variable quantum squeezing is normally observed. This raises the possibility of directly observing the resulting disturbance in the spatial mapping between the local quantum fluctuations of the twin beams. It also suggests that some form of spatial correction, optical or digital, needs to be applied in quantum noise subtraction when doing quantum imaging near field of the OPA.

## ACKNOWLEDGMENTS

J.K. and E.F. acknowledge doctoral training funding from EPSRC.